\documentclass[notitlepage,amsmath,amssymb,superscriptaddress,11pt,floatfix,aps,prd,nofootinbib,preprintnumbers]{revtex4-1}
\usepackage{slashed}
\usepackage{graphicx}
\usepackage{xcolor}
\usepackage{dsfont}
\usepackage[pdftex,final]{hyperref}
\pdfoutput=1

\newcommand{\be}{\begin{equation}}
\newcommand{\ee}{\end{equation}}
\def\beq{\begin{equation}}
\def\eeq{\end{equation}}
\def\bea{\arraycolsep .1em \begin{eqnarray}}
\def\eea{\end{eqnarray}}
\def\s0#1#2{\mbox{\small{$ \frac{#1}{#2} $}}}
\def\0#1#2{\frac{#1}{#2}}
\def\eq#1{(\ref{#1})}
\def\step{\\[-1.5ex]}

\renewcommand{\d}{\mathrm{d}}
\newcommand{\psib}{\bar{\psi}}
\newcommand{\la}[1]{\lambda_{#1}}

\begin{document}
\preprint{CERN-TH-2024-077}

\title{Fermion Mass Generation without  Symmetry Breaking}

\author{Charlie Cresswell-Hogg}
\email{c.cresswell-hogg@sussex.ac.uk}
\affiliation{Department of Physics and Astronomy, University of Sussex, Brighton, BN1 9QH, U.K.}

\author{Daniel F.~Litim}
\email{d.litim@sussex.ac.uk}
\affiliation{Department of Physics and Astronomy, University of Sussex, Brighton, BN1 9QH, U.K.}
\affiliation{Theoretical Physics Department, CERN, 1211 Geneva 23, Switzerland}

\begin{abstract}
We study the generation of fermion mass in a context where interactions break a discrete chiral symmetry. Then, fermion mass is not protected by a symmetry, no symmetry is broken by the generation of mass, and a vanishing mass no longer enhances a symmetry. We elaborate these scenarios for template fermionic and Yukawa theories in three dimensions where mass can be generated either by fluctuations, strong dynamics, or  the vacuum expectation value of a scalar field. We find that fluctuation-induced contributions to fermion mass are parametrically suppressed in the number of fermion flavours $N$. The generation of fermion mass then takes the form of a rapid crossover which turns into a second order quantum phase transition for large $N$, much like in settings with fundamental chiral symmetry. We further discuss theories where fermion mass can be generated spontaneously without breaking any symmetry other than scale symmetry. Implications of our findings are discussed.
\end{abstract}

\maketitle

\tableofcontents

\newpage

\section{\bf Introduction}

Massless Dirac fermions can arise in a large variety of systems ranging from   high-energy  physics to condensed matter.  Dirac fermions can also  acquire a mass, which oftentimes entails the breaking of a discrete or continuous global symmetry such as chiral symmetry or parity. Understanding the mechanisms by which fermions  acquire a mass, dynamically, with or without the breaking of a symmetry, continues to be of interest. Standard
 mechanisms  entail chiral symmetry breaking 
and often  either involve  classically irrelevant  four-fermion interactions and strong dynamics  \cite{Gross:1974jv,Rosenstein:1990nm},
or Yukawa interactions alongside  the condensation of a scalar field.
The  possibility of symmetric mass generation, which refers to settings where fermions become massive, but {\it without} breaking a discrete or continuous chiral symmetry, has received renewed attention  recently. There is a  trail of works in the context of lattice field theory with the aim of giving a mass to fermion doublers in the construction of chiral lattice gauge theories \cite{Creutz:1996xc,
Hasenfratz:1988vc,
Hasenfratz:1989jr,
Lee:1989xq,
Ayyar:2014eua,
Ayyar:2015lrd,
Catterall:2015zua,
Ayyar:2016lxq,
Catterall:2016dzf,
Catterall:2020fep,
Butt:2021brl,
Butt:2021koj}. Symmetric mass generation has also been considered  in the context of   gauge dynamics in the continuum \cite{Tong:2021phe}, and in numerous models of  condensed matter physics, where it relates to novel types of quantum critical points  and topological phase transitions \cite{You:2017ltx,
You:2017mkc,
Xu:2021ztz,
Zeng:2022grc,
Wang:2022ucy,
Lu:2022qtc,
Guo:2023rnz, 
Liu:2023msa}.\step

In this paper, we investigate mass generation in settings where  a discrete chiral symmetry has been made redundant by interactions.  Then, fermion mass  is  no longer protected by a  symmetry,   no discrete symmetry is broken by the  generation of mass, and a vanishing   mass  no longer enhances a symmetry. The aim will be to clarify whether and how effects from chirally-odd interactions percolate and impact upon the generation of mass.  Given that  no discrete or continuous chiral symmetry can   be broken  by the generation of mass, as, per construction, no such symmetries are available, our setup may be viewed as  ``the other side''  of symmetric mass generation.\step

With this in mind, we consider theories with Dirac fermions in three dimensions, covering  purely fermionic theories with Gross-Neveu-type interactions, and   theories with an additional scalar field and Yukawa interactions.  In the presence of fundamental chiral symmetry,  these models showcase  the basic  mechanisms for chiral symmetry breaking and the dynamical generation of fermion mass, which we review. We then move on to   the generation of fermion mass in settings where  chiral symmetry is absent due to  classically irrelevant six-fermion interactions $\sim (\bar\psi\psi)^3$ \cite{Cresswell-Hogg:2022lgg,Cresswell-Hogg:2022lez}, or due to  classically relevant cubic scalar self-interactions $\sim \phi^3$ alongside Yukawa interactions $\sim \phi \bar\psi\psi$ \cite{Cresswell-Hogg:2023hdg}. We are chiefly  interested in understanding how fermion  mass generation through strong dynamics or   the vacuum expectation value of a scalar field competes with mass generation   by fluctuations, which has become available now that fermion mass is no longer protected by a symmetry, and how this competition is influenced by field multiplicities.\step

We are further interested in scenarios where fermion mass is generated spontaneously, following the breakdown of scale symmetry at a quantum critical point \cite{Cresswell-Hogg:2022lgg,Cresswell-Hogg:2022lez,Cresswell-Hogg:2023hdg}. This idea  has seen previous incarnations  in various scalar, supersymmetric, and Chern-Simons theories \cite{Bardeen:1983rv,
David:1984we,
Bardeen:1984dx,
David:1985zz,
Eyal:1996da,
Litim:2011bf,Heilmann:2012yf, Aharony:2012ns,Bardeen:2014paa,Moshe:2014bja,Litim:2017cnl,Litim:2018pxe,Sakhi:2019rfj,Fleming:2020qqx}.
Most notably, fermion mass may take any value without being  determined by  fundamental parameters of the theory.  One also expects to find a light or massless dilaton for sufficiently large $N$ \cite{Omid:2016jve,Litim:2017cnl,Semenoff:2017ptn,Fleming:2020qqx,Semenoff:2024prf}.
 Here, we highlight that for this scenario to be operative, the breaking of chiral symmetry through interactions is a necessary prerequisite to subsequently enable scale symmetry breaking  and the   spontaneous generation of mass. Therefore, fermion mass generation proceeds  without the breaking of a discrete chiral symmetry.
\step

To achieve our results, we employ  functional renormalisation. This continuum method is based on a momentum cutoff and  allows for a continuous interpolation between  microscopic theories at short distances and  quantum effective actions in the IR, without being tied to weak coupling~\cite{Polchinski:1983gv,Wetterich:1992yh,Ellwanger:1993mw,Morris:1993qb}. A virtue of functional renormalisation  in combination with a  large $N$ limit is that closed analytical flow equations can be found \cite{Litim:2000ci,Litim:2001up}, and  that all local potential interactions, in particular   mass terms and interaction vertices at vanishing momenta, can be determined exactly,  without any approximations \cite{DAttanasio:1997yph,Cresswell-Hogg:2022lgg}. This offers a promising starting point to investigate strongly coupled quantum field theories. \step

The manuscript is organised as follows. In Sec.~\ref{Sec:6F}, we investigate  mass generation in purely fermionic theories and contrast the standard scenario of mass generation via strong dynamics with settings where  chiral symmetry is absent due to  interactions, both for finite and infinitely many fermion flavours $N$. Similarly, in Sec.~\ref{Sec:3B}, we investigate mass generation in scalar-Yukawa theories, and contrast scenarios with and without fundamental chiral symmetry. In Sec.~\ref{Sec:FP}, we concentrate on  the spontaneous generation of mass at quantum critical points, which becomes available due to exactly marginal operators.  In Sec.~\ref{Sec:C} we provide a discussion and  some conclusions.

\section{\bf Fermion Mass from Strong Interactions}\label{Sec:6F}

\subsection{Chiral symmetry breaking}
To begin, we recall the classic example of  mass generation through dynamical breaking of chiral symmetry, exemplified by the seminal Gross-Neveu theory \cite{Gross:1974jv}   with $N$ flavours of massless Dirac fermions $\psi_i$ coupled through a four-fermion (4F) interaction  and with fundamental action
\be\label{eq:classical_action4F}
S_{\rm f} = \! \!\int_x  
\Big\{ \psib_i \slashed \partial \psi_i + \frac G2 ( \psib_i \psi_i )^2  \Big\}.
\ee
We mostly consider theories in three dimensions, and take the fermion fields $\psi_i$ to be four-component Dirac fermions. The theory is manifestly  invariant under the discrete symmetry
\be\label{eq:discrete_symmetry}
\psi \to \gamma^5 \psi, \quad \psib \to - \psib \gamma^5\,,\quad \psib \psi\to -\psib \psi\,,
\ee
where $\gamma^5$ refers to an element of the Clifford algebra with $( \gamma^5 )^2 = \mathds{1}$, which corresponds to a three-dimensional analogue of chiral symmetry. Parity can play an equivalent role \cite{ZinnJustin:1991yn,Cresswell-Hogg:2023hdg}.
The theory may be taken as an effective theory at some short distance scale $\Lambda$ to describe phenomena at scales $k \ll \Lambda$. 
Interestingly, and even though the theory is perturbatively non-renormalisable by power counting in $2<d<4$ dimensions,  it is non-perturbatively renormalisable due to the existence of an interacting ultraviolet  fixed point \cite{Wilson:1972cf,Parisi:1975im,Gawedzki:1985ed,Gawedzki:1985jn,Rosenstein:1988pt,deCalan:1991km,Hands:1992be}. 
The four-fermion (4F) coupling $G$ becomes a relevant interaction while all higher order chirally-invariant interactions remain irrelevant. The fixed point guarantees that the effective theory scale $\Lambda$ may actually be  removed $(\Lambda\to\infty)$. The discrete symmetry \eq{eq:discrete_symmetry} further entails that the theory is fundamentally massless, although mass can be generated dynamically.\step

For our purposes, it is convenient to investigate 
the theory~\eqref{eq:classical_action4F} non-perturbatively using functional renormalisation~\cite{Polchinski:1983gv,Wetterich:1992yh,Ellwanger:1993mw,Morris:1993qb} based on the successive integrating-out of momentum modes from a path-integral representation of quantum field theory. The scale-dependence of the corresponding quantum effective action $\Gamma_k$ with respect to the RG momentum scale $k$ is given by  an exact identity \cite{Wetterich:1992yh}
\be
\label{eq:wetterich}
\partial_t \Gamma_k = \tfrac{1}{2} {\rm STr} \left\{ \big[ \Gamma_k^{(2)} + R_k \big]^{-1} \cdot \partial_t R_k \right\}\,,
\ee
with $t=\ln k$,  and $\Gamma_k^{(2)}$  denoting the matrix of second functional derivatives, and the supertrace STr a sum over all momenta and fields. 
The flow \eq{eq:wetterich} interpolates between the microscopic action at short distances $(k\to \Lambda)$ and the full quantum effective action in the infrared $(k\to 0)$, and the infrared cutoff function $R_k(q)$ can be chosen freely within a few constraints~\cite{Litim:2000ci,Litim:2001up,Litim:2001fd}. 
Further, the integrated flow  reproduces standard perturbation theory, it reduces to the exact Callan-Symanzik equation for a mass-like regulator term, and  relates to Polchinski's  RG (based on an UV cutoff) \cite{Polchinski:1983gv}  by a duality transform \cite{DAttanasio:1997yph,Litim:2018pxe}. \step

In the present setting, we consider quantum effective actions with standard kinetic term and general scale-dependent interactions $V_k ( \psib \psi )$ and neglect higher order derivative interactions. This local potential approximation becomes exact in the large-$N$ limit \cite{DAttanasio:1997yph,Cresswell-Hogg:2022lgg,Cresswell-Hogg:2024}. 
It is useful to work in terms of the dimensionless renormalised variables $z = Z_\psi \psib \psi/k^2$ and $v ( z ) = V_k ( \psib \psi )/k^3$. The RG flow for the function $v'$ then takes the form \cite{Jakovac:2013jua,Cresswell-Hogg:2022lgg,Cresswell-Hogg:2022lez}
\be
\label{eq:full_flow_threshold}
\partial_t v = - 3 v + 2 z v' -  \frac32 \int_0^\infty \d y \ y^{\frac{5}{2}} \Bigg\{ \frac{(1+\s01{4 N})\big[ 1+ r (y) \big] \, (-2r')}{y(1+r)^2 + (v')^2}
-\frac{\s01{4 N} \big[ 1+ r (y) \big]  (-2r' )}{y(1+r)^2 +  (v')^2+ 2 z v'v'' } \Bigg\}\,,
\ee
where primes denote partial differentiation with respect to $z$. The  first two terms on the right-hand side account for the classical scaling of the fields and the potential. 
The second set of terms under the momentum integration  account for quantum corrections, with $y=q^2/k^2$ the dimensionless loop momentum, $R_k(q)=\slashed q\, r(y)$ the momentum cutoff,  and $r(y)$ the cutoff shape function  \cite{Litim:2000ci,Litim:2001up}, which we have left unspecified for now. Below, we use an optimised fermionic regulator $r_{\rm opt} ( y ) = ({1}/{\sqrt{y}} - 1 )  \theta \left( 1 - y \right)$~\cite{Litim:2000ci,Litim:2001up,Litim:2001fd,Litim:2002cf} to find simple  expressions, even though key results do not depend on this choice. Finally, we have  scaled an overall normalisation factor into the fields  in accord with na\"ive dimensional analysis  \cite{Weinberg:1978kz}, also enabling the large-$N$ limit.\step

We may expand the function $v$ as  $v ( z ) = \sum_{n = 1}^\infty \la{2n\rm F} \,z^n/{n!}$ in terms of  the $2n$-fermion (2nF) self-interactions $\la{2n\rm F}$. The RG flows of couplings
$\partial_t\la{2n\rm F}\equiv \partial^n_z(\partial_t v)|_{z=0}$  then follow by projection  from \eq{eq:full_flow_threshold}. 
Chiral symmetry imposes that all $\la{2n\rm F}|_{n={\rm odd}}=0$. 
Notice that the functional RG flow is invariant under chiral symmetry provided the initial action and the chosen momentum cutoff respect chiral symmetry. Hence, the infinitesimal integrating-out of momentum modes from the path integral will generate higher order chirally even fermion self-interactions in the effective action, but no chirally odd ones.\step

\begin{figure}[t]
\includegraphics[width=.8\linewidth]{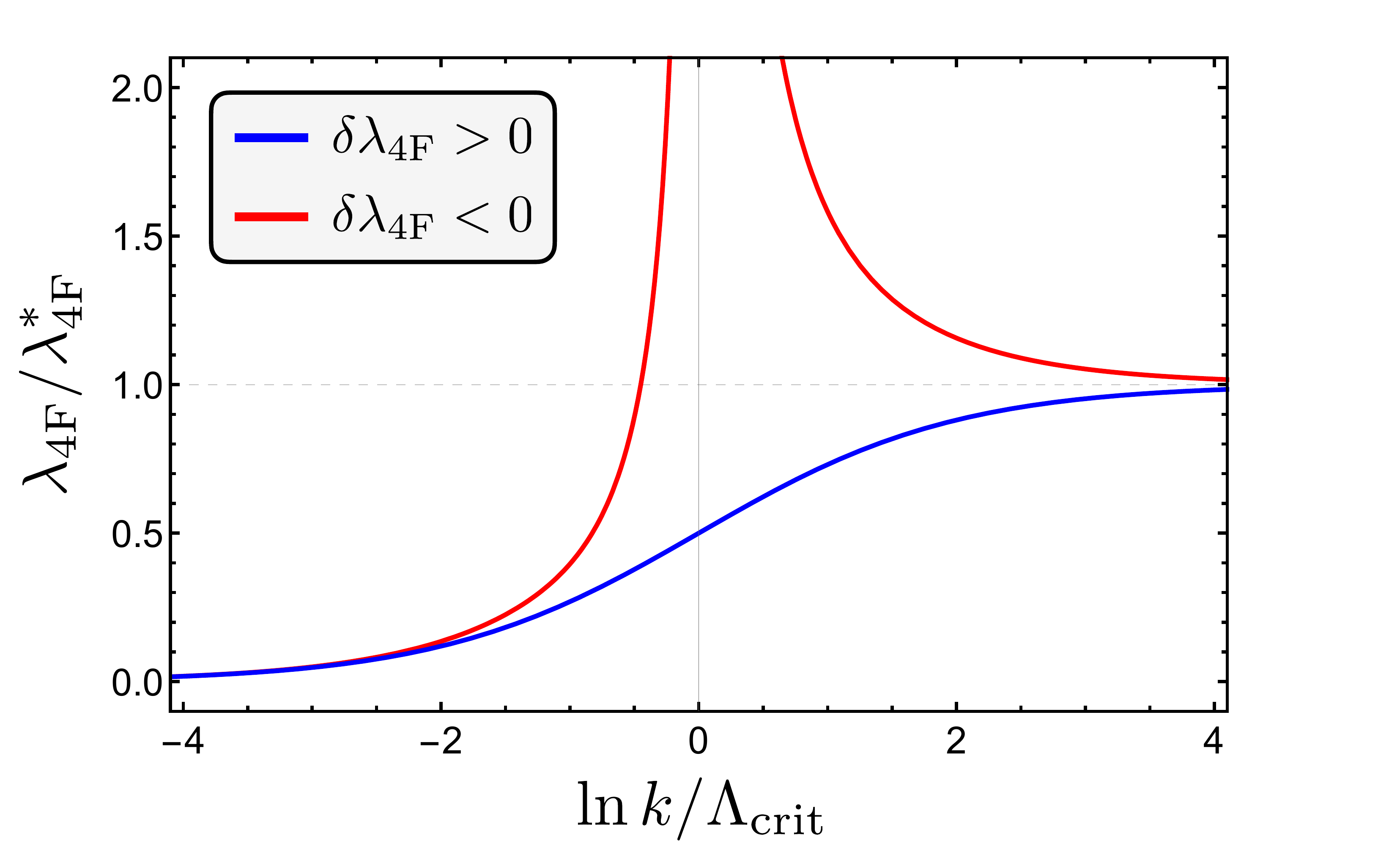}
\caption{Shown is the running 4F coupling \eqref{sol2} along UV-IR connecting trajectories. For $\delta\lambda_{4\rm F}(\Lambda) > 0$  (blue curve),  chiral symmetry remains unbroken, and the 4F coupling displays a smooth cross-over  from UV scaling to IR scaling at the RG-invariant scale $\Lambda_{\rm crit}$. For $\delta\lambda_{4\rm F}(\Lambda)<0$  (red curve), the 4F coupling diverges at $\Lambda_{\rm crit}$,  triggering chiral symmetry breaking  and the generation of fermion mass. In either case, the dimensionful four-fermion coupling $G(k)\equiv \lambda_{4\rm F}/k$ reaches a finite value in the IR $(k\to 0)$.}
\label{fig:4F}
\end{figure}

Next, we are interested in the generation of mass.
Mass can be inserted  ``by hand'', at the expense of breaking chiral symmetry at the high scale via an explicit mass term $ \propto M_F\bar\psi\psi\neq 0$. The RG flow of the dimensionless mass term $m_F\equiv M_F/k$  can be extracted from the projection $\partial_z 
\eq{eq:full_flow_threshold}|_{z=0}$ and reads
\beq\label{beta1}
\partial_t m_F = - m_F \left[1-  6\la{4\rm F} \left(1-\01{4N}\right)
\int_0^\infty\!\! \d y\,  \frac{ -y^{5/2}(1+ r) \, r' (y)}{(y(1+r)^2 + m_F^2)^2}\right] \,.
\eeq
The  first term on the right-hand side accounts for the classical scaling of the mass, and
the second set of terms  for quantum corrections.
Most importantly, at large $N$, the flow \eq{beta1} vanishes identically for vanishing mass, irrespective of the regularisation and the shape of the Wilsonian momentum cutoff function $r(y)$, 
\be\label{mF=0}
\partial_t m_F\big|_{m_F=0}=0\,.
\ee 
This is very different from e.g.~the Higgs mass in the Standard Model, which is switched on by quantum fluctuations, leading to the infamous hierarchy problem. The result \eqref{mF=0} can also be understood as a consequence of chiral symmetry. In fact, the RG flow and the microscopic effective action at the high scale $k=\Lambda$ are chirally symmetric, hence a chirally-odd operator such as the mass cannot be generated by integrating-out infinitesimal momentum shells from the path integral representation of the theory. 
We conclude that fermion mass  is technically natural  in the sense of `t Hooft \cite{tHooft:1979rat} in that it cannot be switched on by fluctuations alone, and that the symmetry is enhanced as soon as $m_F=0$.
\step

Next, we turn to strong coupling effects. A salient feature of the theory relies on the existence of a non-perturbative fixed point in the dimensionless and chirally-even 4F coupling $\lambda_{4 \rm F}\equiv G \,k$,  see \eq{eq:classical_action4F}.  
From its beta function 
\beq\label{beta2}
\partial_t\la{4\rm F}=  \la{4\rm F} + 2\left(1-\frac{1}{2N}\right) \la{4\rm F}^2
\eeq
we observe a free IR fixed point $\la{4\rm F}=0$ and an 
interacting  4F fixed point $\lambda_{4\rm F,*}=-\frac{N}{2N-1}$ for any $N\ge 1$, which is at the root for the non-perturbative renormalisability of the theory. Away from its fixed points, the 4F coupling runs according to
\beq\label{sol2}
\lambda_{\rm 4F}(t)=
\lambda_{\rm 4F}(\Lambda)\,
\frac{k}{\Lambda}
\left[1+ \frac{\lambda_{4\rm F}(\Lambda)}{\lambda_{4\rm F,*}}\left(\frac{k}{\Lambda} -1\right)\right]^{-1}\,. 
\eeq
From the viewpoint of mass generation, what's important with \eq{beta2} and \eq{sol2} is that the fixed point is also responsible for a quantum phase transition between a massless and a massive phase of the theory. Introducing the microscopic parameter $\delta\lambda_{4\rm F}(\Lambda)\equiv \lambda_{4\rm F}(\Lambda)-\lambda_{4\rm F,*}$ with $\Lambda$ the high scale (which we may send to infinity), we observe that the theory remains strictly massless and approaches the free Gaussian limit in the IR for any $\delta\lambda_{4\rm F}>0$, with $\lambda_{4\rm F}$ and all chirally-even couplings smoothly interpolating form the UV fixed point to the free IR fixed point, and $M_F=0$.  The crossover between the UV and IR fixed points occurs at a characteristic energy scale $\Lambda_{\rm crit}$ induced by dimensional transmutation, 
\beq\label{crit}
\Lambda_{\rm crit}=\left|\frac{\Lambda\,\delta\lambda_{4\rm F}(\Lambda)}{\lambda_{4\rm F,*}+\delta\lambda_{4\rm F}(\Lambda)}\right|\,,
\eeq
which, moreover, is an RG invariant ($\frac{d}{d\Lambda}\Lambda_{\rm crit}=0$), much like $\Lambda_{\rm QCD}$ in the theory of strong nuclear interactions. On the other hand, for $\delta\lambda_{4\rm F}<0$,  the 4F interactions become very strong around the scale $\Lambda_{\rm crit}$. Even though the functional flow \eq{eq:wetterich} for actions \eq{eq:classical_action4F} commutes with chiral symmetry \eq{eq:discrete_symmetry},  chiral symmetry breaking is enabled through the occurrence of a cusp-like non-analyticity in the function $V_k ( \psib \psi )$ at vanishing field \cite{Aoki:2014ola}. The non-analyticities  responsible for chiral symmetry breaking occur once $k < \Lambda_{\rm crit}$ and  develop into a cusp for the effective potential in the limit where all quantum fluctuations are integrated out  $(k\to 0)$. In terms of the local 4F coupling \eq{sol2}, the onset of chiral symmetry breaking  is signalled by a Landau-type pole in $\la{4\rm F}$ at the scale $\Lambda_{\rm crit}$ for $\delta\lambda_{4\rm F}<0$. 
The result \eq{mF=0} is thereby circumnavigated, and superseded by the dynamical generation of mass. The mechanism further implies that chiral symmetry is broken maximally, in the sense that not only a fermion mass but all  chirally-odd $2n$F fermion interactions are  switched on in its wake. The running 4F coupling \eqref{sol2} in both the chirally symmetric and dynamically broken phases is illustrated in Fig.~\ref{fig:4F}.
\step

\begin{figure}[t]
\includegraphics[width=.8\linewidth]{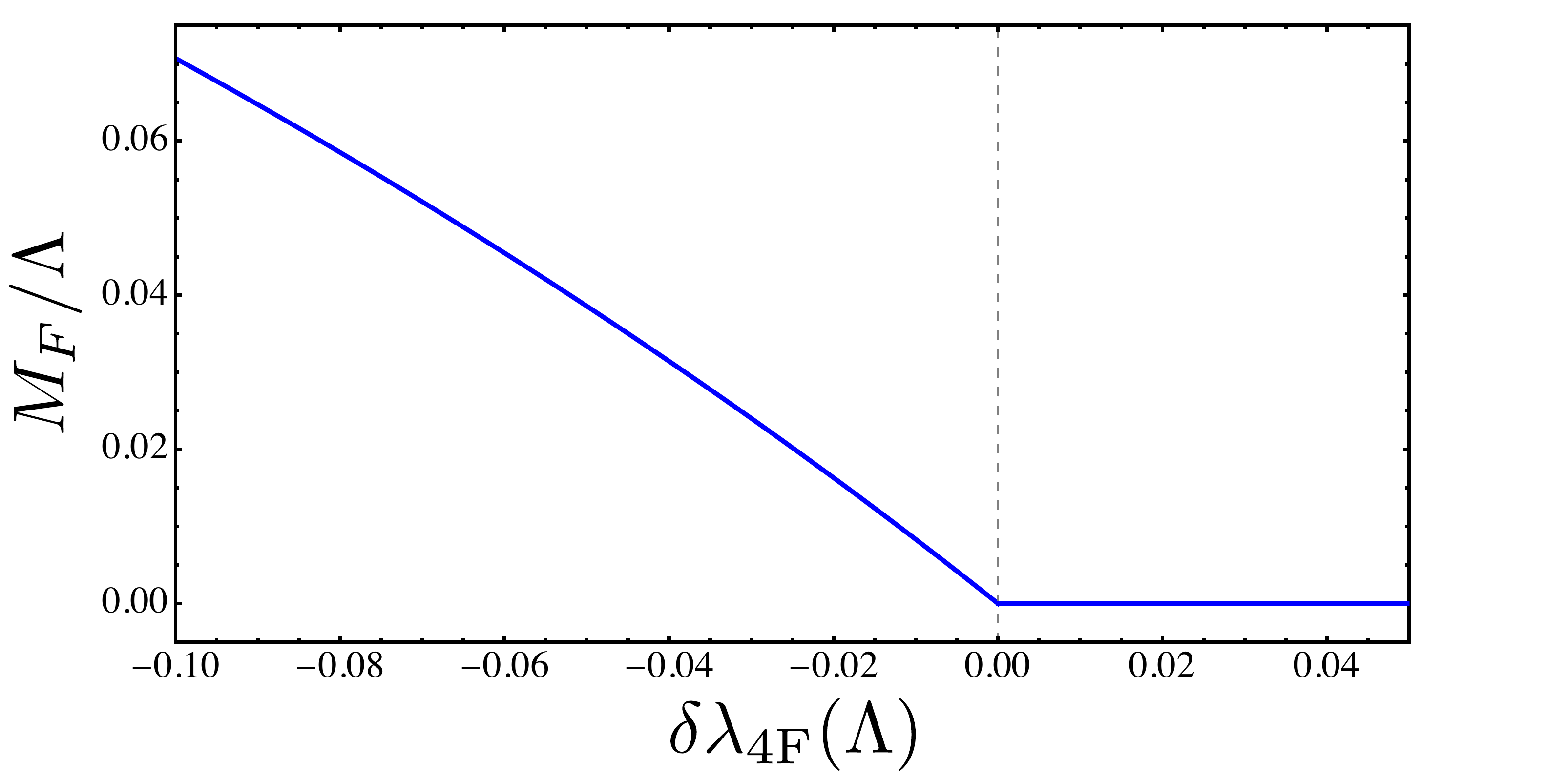}
\caption{Shown is the dynamical generation of fermion mass $M$ through a quantum phase transition in terms of the parameter $\delta\lambda_{4,\rm F}(\Lambda)$. The phase transition dynamically breaks chiral symmetry $(\delta\lambda_{4\rm F}(\Lambda)<0$).}
\label{4FGN}
\end{figure}

Fig.~\ref{4FGN} illustrates the generation of mass quantitatively.\footnote{For the plot, we have used the large-$N$ limit. Similar results arise at finite $N$.}  Here, we have taken an effective theory point of view with $\Lambda$ denoting  a (finite) high scale. As expected, we observe that chiral symmetry is maintained (broken) in the IR provided that  $\delta\la{4\rm F}>0$ ($\delta\la{4\rm F}<0$). We also notice that the  fermion mass in the symmetry broken phase grows with growing $|\delta\la{4\rm F}|$. We may also take the UV limit $\Lambda\to\infty$ to extract universal exponents of the underlying fixed point. With $0<-\delta\lambda_{4\rm F}(\Lambda)\ll 1$ denoting the distance from the 4F quantum critical point ($\delta\la{4\rm F}=0$) in the phase with chiral symmetry breaking, we  extract the scaling exponent $\nu$ of the quantum phase transition 
\beq\label{nu}
M_F \propto \Lambda_{\rm crit}
\propto |\delta\lambda_{4\rm F}|^\nu\quad\text{with}\quad \nu=1\,,
\eeq
by using \eq{beta2} and \eq{crit}.\footnote{The result is exact at infinite $N$, and receives small finite-$N$ corrections from the anomalous dimension and higher order terms in the derivative expansion.}
This is the well-known standard scenario of  fermion mass generation via the dynamical breaking of chiral symmetry, which takes the form of a quantum phase transition in the parameter $\delta\lambda_{4\rm F}$.

\subsection{Symmetric mass generation}

We are now in a position to discuss  fermion mass generation in  settings where chiral symmetry is absent microscopically due to classically irrelevant interactions such as odd powers in $( \psib_i \psi_i )$. The leading one is the six-fermion (6F) interaction $H$,
\be\label{eq:classical_action}
S_{\rm f} = \! \!\int_x  
\Big\{ \psib_i (\slashed \partial +M_F)\psi_i + \frac G2 ( \psib_i \psi_i )^2 + \frac{H}{3!}  ( \psib_i \psi_i )^3 +\cdots \Big\}\,,
\ee
and possibly more. As before, the theory may be taken as an effective theory at some short distance scale $\Lambda$ to describe phenomena at scales $k \ll \Lambda$ \cite{Cresswell-Hogg:2022lgg,Cresswell-Hogg:2022lez}.  \step

Due to the absence of chiral symmetry, switching-on an explicit mass term $M_F\neq 0$ (corresponding to $\lambda_{2\rm F}(\Lambda)\neq 0)$ does not break any symmetry. It results in a massive theory in the IR, irrespective of how the fermion self-interactions are chosen microscopically. Parametrically, the generation of mass then proceeds as a crossover in $\delta\lambda_{4\rm F}$, also implying that a massless phase does not exist for any $\delta\lambda_{4\rm F}$. \step

Next, we are interested in settings where mass is not inserted ``by hand'', $M_F=0$. 
What is perhaps unexpected but important  to observe is that fermion mass is not generated by fluctuations at large $N$, even though chiral symmetry is absent due to interactions. 
This is acknowledged  by looking at the RG flow for the dimensionless mass term $m_F\equiv M_F/k$, which  continues to be given by \eq{beta1}. The crucial point to observe here is that chirally-odd interactions do not contribute to the flow of the mass, and we conclude that fermion mass   cannot be switched on by fluctuations alone. Notice also that  fermion mass is not technically natural  in the  sense of `t Hooft \cite{tHooft:1979rat}, the reason being that  at $m_F=0$, symmetry is not enhanced (recall that chiral symmetry continues to be absent due to interactions). It also follows that the requirement of  chiral symmetry, which amongst others would dictate that all interactions $\sim ( \psib_i \psi_i )^n$ with $n$ odd must vanish identically, is not necessary  to ensure massless fermions. Instead, the significantly milder constraint $m_F=0$ turns out to be sufficient for the latter \cite{Cresswell-Hogg:2022lgg}. \step

\begin{figure}[t]
\includegraphics[width=.8\linewidth]{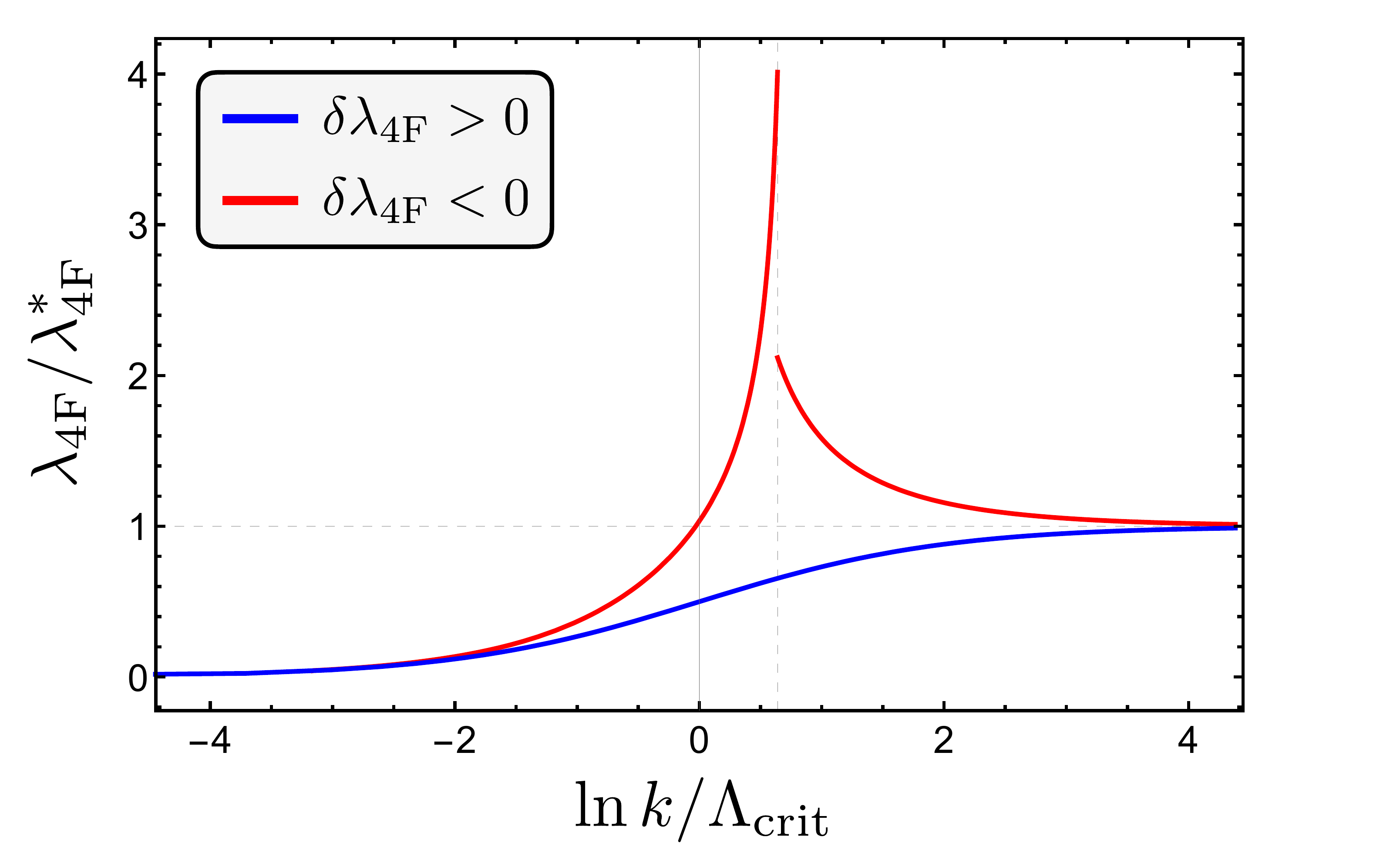}
\caption{Same as Fig.~\ref{fig:4F} but with the addition of 6F interactions  (here $\la{\rm 6F}^* / \la{\rm 6F, crit} = \tfrac45$). For $\delta\lambda_{4\rm F}(\Lambda) > 0$  (blue curve),  the 4F coupling displays a smooth cross-over and the theory remains massless even though chiral symmetry is absent. For $\delta\lambda_{4\rm F}(\Lambda)<0$  (red curve), mass is generated dynamically at the scale $\Lambda_{\rm mass}>\Lambda_{\rm crit}$ (dashed line). Notice that 6F interactions have reduced the Landau pole in $\lambda_{4\rm F}$ to a finite discontinuity. The dimensionful 4F coupling $G(k)\equiv \lambda_{4\rm F}/k$ reaches a finite value in the IR  for any $\delta\lambda_{\rm 4F}$.}
\label{fig:6F}
\end{figure}

We emphasize that the  result \eq{mF=0}, based on \eq{eq:wetterich}, \eq{eq:full_flow_threshold} and \eq{beta1}, is exact in the large-$N$ limit. This implies that the explicit generation of fermion mass at finite $N$ is   suppressed parametrically  in the number of fermion flavours, and at least as $1/N$,
\be\label{mF=1/N}
\partial_t m_F\big|_{m_F=0} = {\cal O}(1/N)\,.
\ee 
Consequently, fermion mass  is parametrically small, and at least as $M_F/\Lambda\propto 1/N$. We note that  inhomogeous contributions $\sim 1/N$ to the flow of the mass \eq{mF=1/N} do not arise through  local potential interactions, nor through  the leading order corrections from anomalous dimensions, sometimes referred to as the LPA$'$ approximation. This pattern is in accord with findings reported in   \cite{Jakovac:2014lqa} for chirally-symmetric settings. We conclude that the explicit  generation of fermion mass  is  delayed in a derivative expansion.
\step

Let us now turn to the leading chirally-odd interactions $ \propto ( \psib_i \psi_i )^3$, with dimensionless coupling $\lambda_{6 \rm F}\equiv H\,k^3$, see \eq{eq:classical_action}. At weak coupling, owing to its large canonical mass dimension, the 6F interaction is strongly  irrelevant. Its beta function, 
$\partial_t\lambda_{6\rm F}=3\lambda_{6\rm F}+\text{quantum corrections}$,
dictates that the coupling scales  rapidly to zero as $\lambda_{6\rm F}\sim (k/\Lambda)^{3}$.  In the vicinity of the strongly-interacting 4F fixed point,   however, it turns out that the 6F interactions are {\it dangerously irrelevant}, the reason being that  they receive large quantum corrections  from interactions.  Using \eq{eq:full_flow_threshold} with  an optimised cutoff $r=r_{\rm opt}$, together with $\la{2\rm F}=0$ and \eq{beta2},  we find
\be\label{betala3}
\partial_t\lambda_{6\rm F}=\frac{3}{4N}\lambda_{6\rm F}+ 6 \,\delta \la{4\rm F}(t)\, \la{6\rm F}\, ,
\ee
where we have dropped terms subleading in $1/N$. At the 4F fixed point ($\delta \la{4\rm F}=0$), we observe that $\lambda_{6\rm F}$ becomes   parametrically marginal in $1/N$ due to quantum fluctuations. Integrating \eq{betala3} together with \eq{beta2} in terms of initial conditions for the 4F and 6F interactions at the high scale gives the explicit solution
\beq\label{la3}
\la{6\rm F}  = \la{6{\rm F},\Lambda} \left(\frac{k}{\Lambda} \right)^\frac{3}{4N} 
\left[1+ \left( 1-\frac{\Lambda}{k}\right)\frac{\delta\lambda_{4{\rm F},\Lambda}}{\lambda_{4\rm F,*}}\right]^{-\frac{6N}{2N+1}}\,.
\eeq
For RG scales above \eq{crit}, $\Lambda\ge k\gg \Lambda_{\rm crit}$, and small $\delta \la{4\rm F}$, the running is dominated by the 4F fixed point. As a consequence, we observe from \eq{la3} that the canonical running of $\la{6\rm F}$  is reduced by quantum effects, leading to a much slower running $\lambda_{6\rm F}\sim (k/\Lambda)^{3/(4N)}$. On the other hand, as soon as the 4F coupling has deviated significantly from its fixed point $(k\ll \Lambda_{\rm crit})$, near-canonical scaling takes over and the running becomes much faster $\lambda_{6\rm F}\sim (k/\Lambda)^{3-3/(4N)}$. In the infinite-$N$ limit, $\lambda_{6 \rm F}$ becomes exactly marginal  \cite{Cresswell-Hogg:2022lgg}, and the  fixed point \eq{beta2}  degenerates into a finite line of fixed points parametrised by the 6F coupling $\lambda_{6\rm F}\in [-\la{6\rm F, crit},\la{6\rm F, crit}]$, which becomes a new fundamentally free parameter of the theory.\step

\begin{figure}[t]
\includegraphics[width=.8\linewidth]{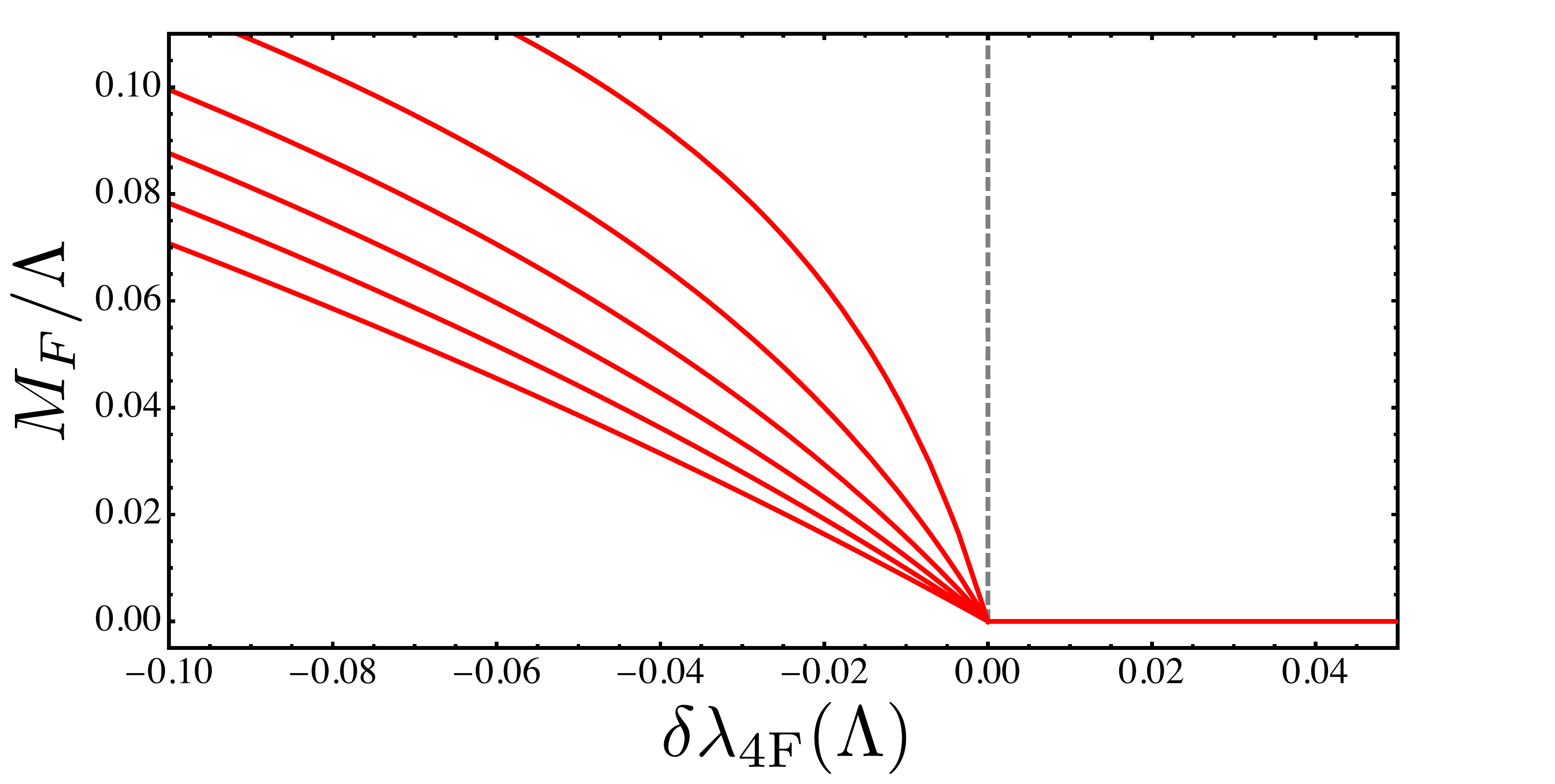}
\caption{Symmetric mass generation in terms of the parameter $\delta\lambda_{4\rm F}(\Lambda)$ and different values for the six fermion coupling $\la{6\rm F}(\Lambda)> 0$ ($\la{6 \rm F}(\Lambda) = \frac{n}{6} \la{6\rm F, crit}$, $n = 0, \dots, 5$ from bottom to top).}
\label{6FGN}
\end{figure}

The large-$N$ result \eq{mF=0} implies that the generation of mass still proceeds through a quantum phase transition with critical exponent \eq{nu} even though chiral symmetry is not observed. For $\delta\la{4\rm F}(\Lambda)>0$, 
we find that all RG trajectories are attracted by the free Gaussian fixed point  in the IR for any 2nF coupling, $\lambda_{2n\rm F}(k\to 0)\to \lambda^*_{2n\rm F}=0$. In its vicinity, all dimensionful couplings $\bar\lambda_{2n\rm F}\equiv\lambda_{2n\rm F}(k) \,k^{3-2n}$  settle at finite values in the IR limit, as long as $\lambda_{2n\rm F}\neq 0$ sufficiently close to the Gaussian. The theory remains strictly massless, however, $\bar\lambda_{2\rm F}\equiv M_F=0$, owing to the result \eq{mF=0}.\footnote{We note that chiral symmetry  could  emerge in the IR where  the Gaussian fixed point is reached \cite{Cresswell-Hogg:2022lgg}. This  requires that all chirally-odd couplings vanish sufficiently close to the Gaussian, $\lambda_{2n\rm F}= 0$ ($n=$~odd), to ensure $\bar\lambda_{2n\rm F}= 0$ in the IR. The latter, however, necessitates chiral symmetry at the high scale.} \step

On the other hand, once $\delta\la{4\rm F}(\Lambda)<0$, the generation of mass relates, once more, to a discontinuity in the function $V'_k$  for vanishing field at the scale $k= \Lambda_{\rm mass}$. Interestingly, however, as soon as $\lambda_{\rm 6F}\neq 0$, all $2n$-fermion coupling $\lambda_{\rm 2nF}$ remain finite across the discontinuity. This is different from what happens for $\lambda_{\rm 6F}=0$, where  $\lambda_{\rm 4F}$  runs through a Landau pole (Fig.~\ref{fig:4F}). In other words, the parity-odd 6F interaction smoothens-out the dynamical generation of mass and reduces the Landau pole in $\lambda_{\rm 4F}$  to a finite discontinuity (Fig.~\ref{fig:6F}). We also observe that while the RG-invariant scale of  fermion mass generation $\Lambda_{\rm mass}$ (which now depends on the UV parameter $\lambda^*_{\rm 6F}$) coincides with the RG invariant scale $\eq{crit}$ in parity-even theories (where $\lambda^*_{\rm 6F}=0$), it comes out increasingly larger $\Lambda_{\rm mass}>\Lambda_{\rm crit}$  with increasing $|\lambda^*_{\rm 6F}|\neq 0$. This result also explains why the 6F coupling does not reach a Landau pole at $\Lambda_{\rm crit}$, even though its explicit running \eqref{la3} would have suggested that it might, because after mass has been generated at $\Lambda_{\rm mass}>\Lambda_{\rm crit}$,  the flow \eqref{la3} is superseded by the corresponding flow in the massive theory with  no Landau poles below $\Lambda_{\rm mass}$.
\step

At  finite $N$,  the  presence of parametrically subleading corrections \eq{mF=1/N} implies that
the generation of mass  proceeds through a  rapid crossover in $\delta\la{4\rm F}(\Lambda)$, rather than a    continuous phase transition. We emphasise, however, that the dynamically-generated part of fermion mass is of order unity (Fig.~\ref{6FGN}), and parametrically larger than any fluctuation-induced part. Therefore, the crossover becomes increasingly rapid in the limit $1/N\to 0$, where it turns into a  second order quantum phase transition.
\step

Fig.~\ref{6FGN} illustrates the generation of mass at large $N$ quantitatively. 
Denoting by $\Lambda$   the high scale and the initial microscopic parameters by $\delta\la{4\rm F}(\Lambda)$ and $\lambda_{\rm 6F}(\Lambda)$, we observe that the theory remains in a massless (massive) phase  in the IR provided that  $\delta\la{4\rm F}>0$ ($\delta\la{4\rm F}<0$). In the massive phase,  the  fermion mass grows with growing $|\delta\la{4\rm F}|$. In the presence of chirally-odd interactions, the growth is additionally enhanced by  6F interactions.\footnote{We recall that the magnitude of the generated fermion mass $| M_F |$ is  invariant under $\lambda_{\rm 6F}\leftrightarrow -\lambda_{\rm 6F}$, and grows with growing $\lambda_{\rm 6F}>0$, ${\rm d} | M_F(\lambda_{\rm 6F}) |/{\rm d}\lambda_{\rm 6F}>0$, for any given $\delta\lambda_{\rm 4F}(\Lambda)<0$.} Ultimately, the  reason for this is  that  both 4F and 6F  interactions become very strong. It is also  noteworthy that the apparent Landau poles around the scale $\Lambda_{\rm crit}$, see \eq{sol2}, \eq{la3} are not reached, the reason being that dynamical mass generation at $\Lambda_{\rm mass}$ already happened prior to it.  Hence, our findings illustrate that dynamical mass generation in fermionic theories can proceed without an underlying Landau pole in the 4F interactions, curtesy of the 6F interactions. Since a discrete chiral symmetry is absent in either phase, no symmetry is broken when fermion mass is generated.

\section{\bf Fermion Mass from Yukawa Interactions}\label{Sec:3B}
\subsection{Chiral symmetry breaking}

We  recall  the mechanism of fermion mass generation via (chiral) symmetry breaking in Yukawa theories. Consider $N$ flavours of four-component Dirac fermions $\psi_i$ coupled to a  real scalar field $\phi$, with classical action \cite{Cresswell-Hogg:2023hdg}
\be\label{eq:SGNY}
S = \! \int \! \d^3 x \left\{ \psib_i \slashed{\partial} \psi_i \! + \tfrac 12 ( \partial \phi )^2 + H  \phi  \, \psib_i \psi_i \! + \tfrac 12 M_s^2 \phi^2
+\s0{1}{4!}\bar\lambda_4\phi^4
+ \s0{1}{6!}\bar\lambda_6\phi^6
\right\} .
\ee
Interactions in these theories are parametrised by the Yukawa coupling $H$ and  the scalar self-interactions such as the quartic or sextic couplings  $\lambda_4$ and $\la{6}$. Besides a global $U(N)$ flavour symmetry, the theory \eqref{eq:SGNY} is  invariant under the discrete transformations
\be\label{eq:discreteSym}
\psi \to \gamma^5 \psi, \ \psib \to -\psib \gamma^5, \ \phi \to -\phi\,,
\ee
which is the bosonised version of \eq{eq:discrete_symmetry}.
Demanding invariance under  \eq{eq:discreteSym} entails that any interactions $\sim \phi^{n} (\psib\psi)^{m}$ with $n+m=$~odd are forbidden. Consequently, a scalar mass term $\sim M^2_s\,\phi^2$ is allowed, but  a fermion mass  $\sim M_F\,\psib\psi$ is not.\step

The theory is perturbatively renormalisable and displays an asymptotically free fixed point in the UV. 
Further, in the IR, the theory displays an interacting fixed point, characterising a quantum phase transition between a symmetric phase where the expectation value of the scalar field vanishes, $\langle \phi\rangle=0$, and a symmetry broken phase where it does not, $\langle \phi\rangle\neq 0$. The  transition between the symmetric $(M_s^2 >0)$ and the symmetry broken phase $(M_s^2 <0)$ is controlled by the scalar mass term, much like in the Ising universality class. Crucially, the ground state thereby can break chiral (parity) symmetry in the scalar sector, and the Yukawa interaction is then responsible for generating a fermion mass proportional to the vev of the scalar field $M_F\sim H\langle \phi\rangle$. This is the standard mechanism of fermion mass generation, whereby (chiral) symmetry breaking  spills over from the scalar sector  into the fermionic sector through Yukawa interactions.\step

To be more explicit, we consider the functional RG flow \eqref{eq:wetterich} for the theory \eqref{eq:SGNY}   in three euclidean dimensions. 
Introducing a dimensionless and renormalised  scalar field   $ \sigma = Z_\phi^{1/2} \, \phi/k^{1/2}$, Yukawa coupling $h =( Z_\phi \,Z_\psi^2)^{-1/2} \,H/k^{1/2}$, and a dimensionless potential $u ( \sigma) =  U_k ( \phi )/k^{3}$ comprising all chirally even polynomial scalar interactions $u=\sum_n\s01{(2n)!}\la{2n} \sigma^{2n}$, we find their RG equations
 \begin{align}
\label{eq:flowu}
\partial_t u &= -3 u + \frac12 \left( 1 + \eta_\phi \right) \sigma \partial_\sigma u - \frac32 \int_0^\infty \!\! \d y \ \frac{y^{\frac52} \left( 1+ r \right) \, (-2r')}{y(1+r)^2 + (h \sigma)^2}
\\
\label{eq:flowh}
\partial_t h^2 &= - \left( 1 - \eta_\phi \right) h^2\,.
\end{align}
The scalar field anomalous dimension reads  $\eta_\phi \equiv h^2/h^2_*$ in terms of the Yukawa coupling and its IR fixed point $h_*$. The fermion anomalous dimension $\eta_\psi$ vanishes at large $N$ and $Z_\psi=1$. The first terms in \eqref{eq:flowu}  arise from classical mass dimensions of the potential and the fields. The last  term   accounts for quantum corrections, written in terms of a momentum integral with $y=q^2/k^2$ the dimensionless loop momentum,  and $r(y)$ the cutoff shape function which we have left unspecified for now  \cite{Litim:2000ci,Litim:2001up}. \step

The theory \eq{eq:SGNY} with \eq{eq:flowu} and \eq{eq:flowh} displays an IR attractive, and non-perturbatively interacting fixed point in the Yukawa sector $\eta_\phi|_*=1$, giving the IR Yukawa fixed point $h^2_*=\s052$ for the optimised cutoff.\footnote{The existence of the fixed point and its scaling dimension are universal and scheme-independent, but its value is non-universal and may take different values for different RG cutoff functions $r(y)$. Also, both solutions $h_*=\pm\sqrt{2/5}$  are equivalent as the sign of the Yukawa coupling is irrelevant. Below we take the positive root without loss of generality.}  It dictates that the scalar field anomalous dimension takes an integer value in the IR, and   induces a fixed point for the entire scalar potential, 
 \begin{align}\label{eq:FPsolutions}
u_* ( \sigma ) &= ( h_* \sigma )^3 \arctan ( h_* \sigma )+ ( h_* \sigma )^2 \,,
\end{align}
normalised to $u_*(0)=0$. To find the expression \eqref{eq:FPsolutions},  we use an optimised fermionic regulator $r_{\rm opt} ( y ) = ({1}/{\sqrt{y}} - 1 )  \theta \left( 1 - y \right)$~\cite{Litim:2000ci,Litim:2001up,Litim:2001fd,Litim:2002cf}, similar expressions are found for other regulator shape functions. The IR fixed point characterises a quantum phase transition between a symmetric phase with $\langle \phi\rangle=0$ and global chiral symmetry, massless fermions, and a massive scalar, and a symmetry broken phase with $\langle \phi\rangle\neq 0$ and dynamically broken chiral symmetry,  massive fermions, and a massive scalar. \step

\begin{figure}[t]
\includegraphics[width=.7\linewidth]{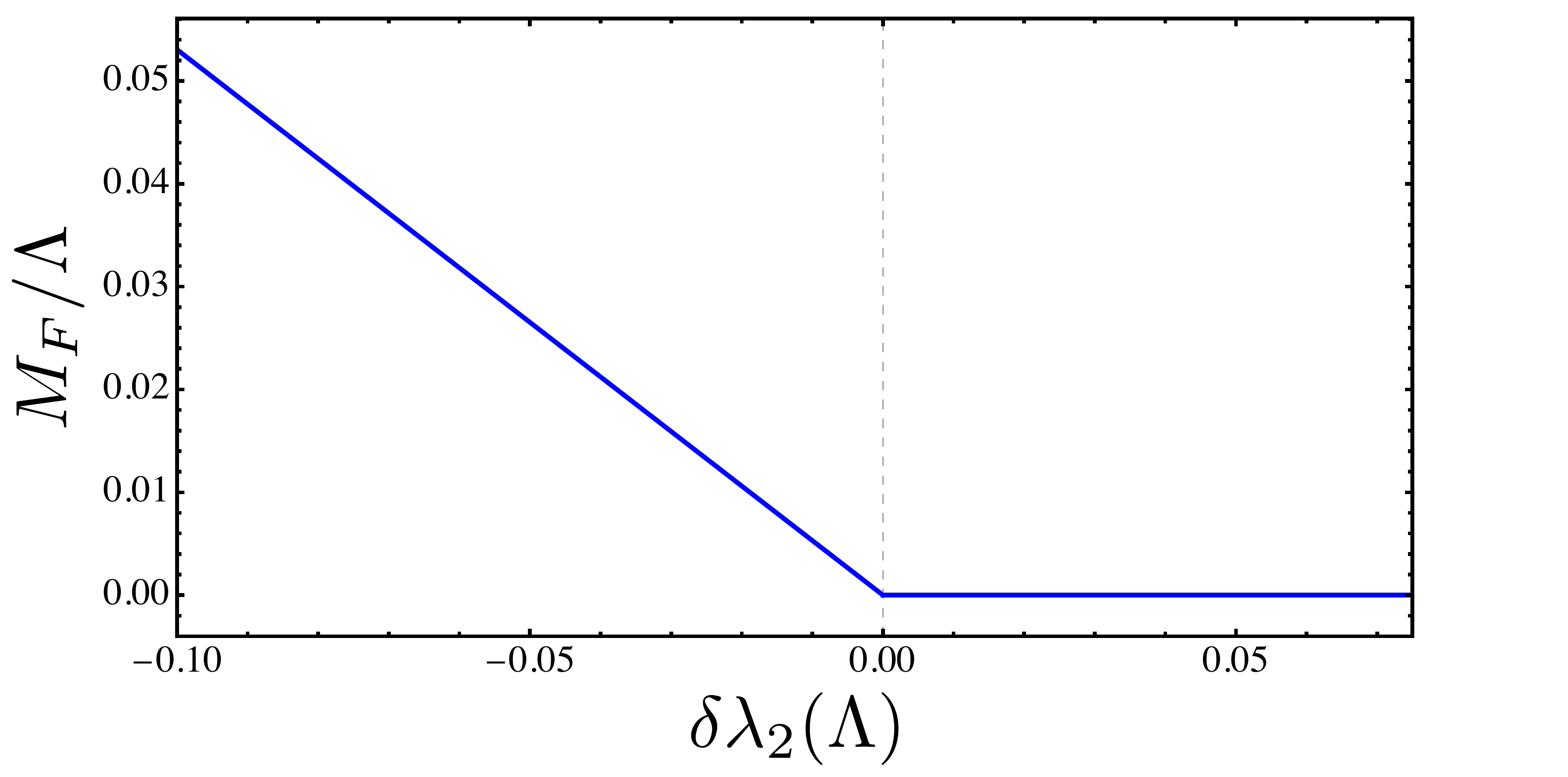}
\caption{Shown is the dynamical generation of fermion mass $M$ through a quantum phase transition in terms of the parameter $\delta\lambda_{2}(\Lambda)$. For $\delta\lambda_{2}(\Lambda)<0$, the phase transition dynamically breaks chiral symmetry, and also switches on chirally-odd  fermion couplings $\sim ( \psib_i \psi_i )^n$ with $n$ odd.}
\label{2BGNY}
\end{figure}

The quantum phase transition is best seen by using  a relevant perturbation in the scalar mass term $\delta\la{2}=\la{2}-\la{2,*}$ to trigger an RG flow away from the fixed point, see Fig.~\ref{2BGNY}. For $\delta\la{2}=0$, the massless theory remains at the fixed point.  For $\delta\la{2}>0$, the theory remains in the symmetric phase, yet the scalars become massive.  For  $\delta\la{2}<0$  a non-trivial vacuum expectation value arises for the scalar field. This breaks  chiral symmetry dynamically, and the fermions become massive via the Yukawa interactions with critical exponent
\beq\label{nuGNY}
M_F\propto |\delta\lambda_2|^\nu\quad\text{with}\quad \nu=1\,.
\eeq
This is the well-known standard scenario of  fermion mass generation via the dynamical breaking of chiral symmetry through Yukawa interactions and a scalar vacuum expectation value.\step

Finally, we are interested in the explicit generation of fermion mass, where taking $M_F \neq 0$  breaks chiral symmetry explicitly.  To that end, we look into the flow for the dimensionless fermion mass $m_F=M_F/k$, which in the limit of many fermion flavours takes the form 
\be\label{eq:mF2}
\partial_t m_F = - m_F \left( 1 + \frac{6 \,h^2}{\la{2}} \int_0^\infty\!\! \d y\,  \frac{ -y^{5/2}(1+ r) \, r' (y)}{(y(1+r)^2 + m_F^2)^2} \right) \,.
\ee
Here $\la{2}=M_s^2/k^{2-\eta_\phi}$ denotes the renormalised  scalar mass squared in units of the RG momentum scale $k$. The first term in \eqref{eq:mF2}  accounts for the canonical mass dimension of $M_F$ and the second one  for quantum corrections. Most importantly, we observe that fermion mass is technically natural in the sense of `t~Hooft \cite{tHooft:1979rat} in that it cannot be switched on by fluctuations, see \eq{mF=0}, and that the distinguished point $m_F =0$ enhances the symmetry to full chiral symmetry. \step

\subsection{Symmetric mass generation}
Next, we turn to settings where chiral symmetry is absent from the outset due to classically relevant interactions, starting with the action
\be\label{eq:SGNY3}
S = \! \int \! \d^3 x \left\{ \psib_i \slashed{\partial} \psi_i \! 
+ \tfrac 12 ( \partial \phi )^2 
+ H  \phi  \, \psib_i \psi_i \! 
+ \tfrac 12 M_s^2 \phi^2
+\s0{1}{3!}\bar\lambda_{3}\phi^3
+\s0{1}{4!}\bar\lambda_4\phi^4
+\s0{1}{6!}\bar\lambda_6\phi^6 
\right\} .
\ee
The main new addition as opposed to \eq{eq:SGNY} are cubic scalar self-interactions $\sim \phi^3$, which break chiral symmetry \eq{eq:discreteSym} at the microscopic level \cite{Cresswell-Hogg:2023hdg}.  In the UV, the cubic interaction corresponds to a relevant perturbation with mass dimension $[\bar\lambda_{3}]=\s032$. The theory continues to be asymptotically free and renormalisable in perturbation theory, and, much like its chirally symmetric counterpart \eq{eq:SGNY}, displays a critical point in the IR. However, given that chiral symmetry is absent from the outset, it is no longer  meaningful to think in terms of symmetric and symmetry broken phases. Instead, we refer to settings with $\langle \phi\rangle\neq 0$ or $\langle \phi\rangle=0$ as the ``ordered'' or ``disordered'' phase, respectively. \step

The RG running of couplings in the theory \eqref{eq:SGNY3} is given by the same set of functional RG equations as before, \eqref{eq:flowu} and \eqref{eq:flowh}, only that the scalar potential now also includes interactions odd under parity.  In particular, for the renormalised dimensionless cubic coupling $\lambda_3=\bar \lambda_3 /(k^{3/2}\,Z_\phi^{3/2})$, we find the RG flow
 \be\label{eq:flow-lambda3}
\partial_t \la{3} = -\tfrac32(1 - h^2/h^2_* )\, \la{3}
\ee
in the large-$N$ limit. Notice that the canonically relevant cubic coupling becomes exactly marginal non-perturbatively, precisely due to the  fixed point in the Yukawa coupling \eqref{eq:flowh} which enforces $h^2 \to h^2_*$ in the IR. As such,  the classically relevant cubic coupling  in the Yukawa theory with beta function \eqref{eq:flow-lambda3} plays a role similar to the  dangerously irrelevant 6F interactions in the fermionic theory \eq{eq:classical_action}  with beta function $\partial_t\lambda_{6\rm F}=+3 (1-\la{4\rm F}/\la{4\rm F,*})\, \la{6\rm F}$. The result \eq{eq:flow-lambda3} has the important implication that  $\la{3}$ becomes a free parameter at $h=h_*$, opening up an entire  line of fixed points with critical potentials 
 \begin{align}\label{eq:FPsolutions3}
u_* ( \sigma ) &= \s016\la{3}\sigma^3+
( h_* \sigma )^3 \arctan ( h_* \sigma )+ ( h_* \sigma )^2 \,,
\end{align}
parametrised by the cubic scalar coupling $\lambda_3$. The latter is allowed to take values in a finite range  $[-\la{3,\rm crit},\la{3,\rm crit}]$. Outside this range, the ground state of the theory becomes unstable, and the fixed point no longer describes a viable physical theory.\step

\begin{figure}[t]
\includegraphics[width=.7\linewidth]{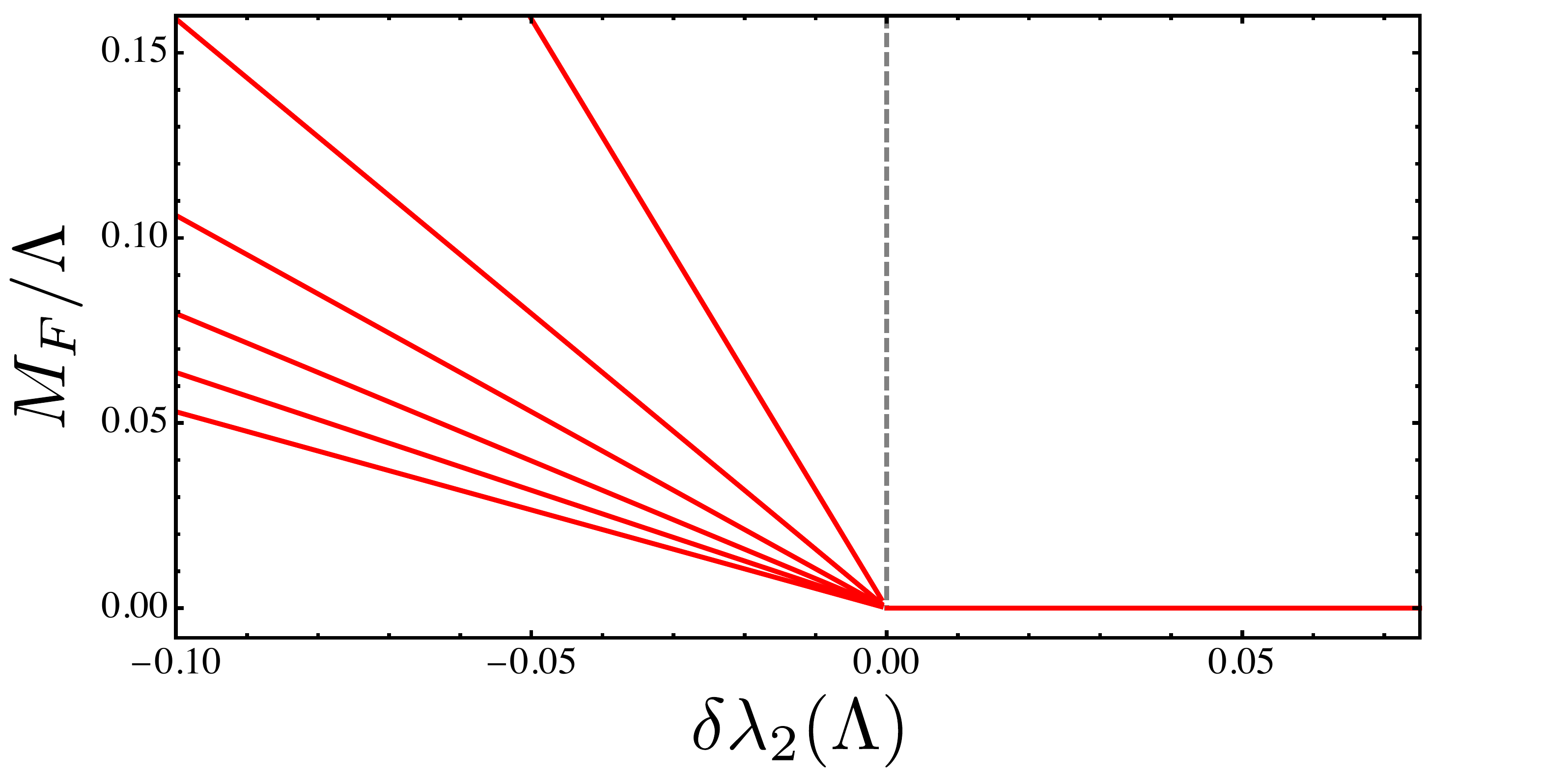}
\caption{Symmetric mass generation in terms of the parameter $\delta\lambda_{2}(\Lambda)$ and various $\la{3}(\Lambda) < 0$ ($\la{3}(\Lambda) = -\frac{n}{6} \la{3, \rm crit}$, $n = 0, \dots, 5$ from bottom to top). }
\label{3BGNY}
\end{figure}

Due to the absence of chiral symmetry, switching-on an explicit fermion mass term $M_F\neq 0$  ``by hand'' does not break any symmetry, and leads to a massive theory in the IR. Parametrically, the generation of fermion mass then proceeds as a crossover in $\delta\lambda_2$, also implying that a massless phase no longer exists for any $\delta\lambda_2$. 
\step

Next, we are interested in whether fermion mass can be generated by fluctuations, starting with $M_F = 0$ microscopically. Starting with  the large-$N$ limit,  we find that the flow for the dimensionless fermion mass $m_F=M_F/k$ continues to be given by \eq{eq:mF2}. Crucially, chirally-odd interactions do not contribute to \eq{eq:mF2}, and  mass is still not switched on by fluctuations even though chiral symmetry is absent.
Rather, the constraint $m_F=0$ is sufficient to protect fermion mass, and  the stronger constraint of full chiral symmetry is not required. Notice, however,  that fermion mass is not technically natural in the  sense of \cite{tHooft:1979rat}, even though its RG flow is proportional to mass itself, the reason being that a vanishing mass does not lead to an enhancement of symmetry.  \step

Consequently, the line of IR fixed points now characterises a quantum phase transition between a disordered phase  $\langle \phi\rangle=0$ with massless fermions and a massive scalar, and an ordered phase  $\langle \phi\rangle\neq 0$ with  massive fermions and a massive scalar. Specifically, using the scalar mass term as a relevant perturbation $\delta\la{2}=\la{2}-\la{2,*}$, the theory remains in the  disordered phase for $\delta\la{2}>0$ where the fermions are massless.  For  $\delta\la{2}<0$  a non-trivial vacuum expectation value arises $\langle \phi\rangle\neq 0$ whereby the scalar becomes massive, and the fermions  become massive via Yukawa interactions. We also note that increasing the value  of the cubic scalar interactions enhances the fermion mass in comparison to its value in the chirally symmetric setting (see Fig.~\ref{3BGNY}).
Close to the transition one finds the   critical scaling
\beq
M_F\propto |\delta\lambda_2|^\nu\quad\text{with}\quad \nu=1\quad (\delta\la{2}<0)\,.
\eeq
By and large, this resembles the  standard scenario of  fermion mass generation via Yukawa interactions except that no global symmetry is broken. Once more, the generation of fermion mass proceeds through a quantum phase transition, curtesy of \eq{eq:mF2}, and without the breaking of a symmetry. \step

Finally, we consider finite $N$ corrections to the above. What changes at finite $N$ is that  fluctuations  induce further Yukawa interaction terms $\sim h_n\,k^{(2-n)/2}\int \phi^n \psib\psi$ beyond those present in \eqref{eq:SGNY3}, and that inhomogeneous $1/N$ corrections to the flow   \eq{eq:mF2} generate mass explicitly. We find
\beq\label{GNY-mF=1/N}
\partial_t m_F\big|_{m_F=0} = \frac{1}{4 N} \left( 1 - \frac{\eta_\phi}{5} \right) \frac{h_1 \la{3} - h_2 \la{2}}{\la{2} \, ( 1+ \la{2} )^2}
\eeq
in three dimensions. Here, $\la{2}=M_s^2/k^2$ denotes the scalar mass in units of the RG scale.  Notice that both the chirally-odd cubic scalar self-interactions $\sim \la{3}$ and the chirally-odd Yukawa coupling $\sim h_2$ contribute to the explicit generation of mass, and that $h_1 \la{3} - h_2 \la{2}\neq 0$ in general. 
We conclude from \eqref{GNY-mF=1/N} that  mass is generated due to fluctuations, and that the effect  is  suppressed parametrically as $1/N$.  In Fig.~\ref{3BGNY}, this corresponds to a parametrically small offset $\Delta (M_F/\Lambda) \sim 1/N$, meaning and that the generation of fermion mass proceeds through  a rapid cross-over, which reduces to a proper quantum phase transition in the limit $1/N\to 0$.

\section{\bf Fermion Mass from Spontaneously Broken  Scale Symmetry}\label{Sec:FP}
In this section, we investigate  models where fermion mass generation is triggered via the  spontaneous breaking of scale symmetry at a strongly interacting  conformal fixed point. \step

The possibility that mass can be generated  at a quantum critical point  through the  spontaneous breaking of scale invariance has first been exemplified in the context of large $N$ scalar $(\phi^2)^3_{\rm 3d}$  theories \cite{Bardeen:1983rv}, see also \cite{David:1984we,David:1985zz,Litim:2017cnl,Litim:2018pxe,Fleming:2020qqx}. Subsequently, this phenomenon has been observed  in multicritical bosonic theories \cite{Eyal:1996da}, in supersymmetric  theories \cite{Bardeen:1984dx,Litim:2011bf,Heilmann:2012yf}, in 3d models where bosons or fermions are coupled to a topological Chern-Simons term \cite{Aharony:2012ns,Bardeen:2014paa,Moshe:2014bja,Sakhi:2019rfj}, and, more recently, in  large $N$ fermionic $(\bar\psi\psi)^3_{\rm 3d}$  theories  \cite{Cresswell-Hogg:2022lgg,Cresswell-Hogg:2022lez} and in closely-related  3d Yukawa theories \cite{Cresswell-Hogg:2023hdg}.
A common feature  is that the breaking of an exact or approximate scale symmetry leads to a massless or light Nambu-Goldstone boson, the dilaton \cite{Omid:2016jve,Litim:2017cnl,Semenoff:2017ptn,Fleming:2020qqx,Semenoff:2024prf}. In the context of particle physics, it has been argued that spontaneously broken scale symmetry may be operative at the lower end of the Banks-Zaks conformal window of QCD \cite{DelDebbio:2021xwu,Zwicky:2023bzk,Zwicky:2023krx}, and that the Higgs particle may arise as a light dilaton in suitable extensions of the Standard Model \cite{Goldberger:2008zz,Bellazzini:2012vz,Csaki:2015hcd}.

\subsection{Critical Gross-Neveu}

We  consider 3d Gross-Neveu theories \eqref{eq:classical_action}  in the  infinite-$N$ limit \cite{Cresswell-Hogg:2022lgg,Cresswell-Hogg:2022lez}. In this limit, the local potential approximation becomes exact and exactly solvable \cite{DAttanasio:1997yph,Cresswell-Hogg:2022lgg}, which makes it an ideal starting point for our purposes.\footnote{For functional RG flows in a local potential approximation, the difference between finite and infinite $N$  theories evidences itself through a reduction of the flow from a second to a first order partial differential equation, whence solutions may be qualitatively different. In the context  of  scalar field theories, we  refer to e.g.~\cite{Litim:2016hlb,Juttner:2017cpr,Litim:2017cnl,Litim:2018pxe,Fleming:2020qqx,Yabunaka:2021fow} for aspects of convergence for $N \to \infty$, uniqueness, and potential ambiguities.} As can be seen from \eq{beta2}, \eq{betala3}, the theory displays an exact line of UV conformal critical points characterised by the value of the relevant 4F coupling $\la{4\rm F,*}$, and the exactly marginal 6F coupling $\la{6\rm F}$ with
\beq \partial_t \lambda_{\rm 6F}\equiv 0\,,\eeq
which serves as a free parameter. At the fixed point, the theory is scale-invariant and massless. Once $\la{6\rm F}\neq 0$, chiral symmetry is absent fundamentally. Then the line of fixed point effective actions are described by a classical kinetic term and a local interaction potential $v_*(z)$ involving arbitrarily high interaction monomials. The latter  is determined implicitly by the relation
\beq\label{eq:FP6F}
z=-2 v_*' +4 (v_*')^2\left[\la{6\rm F}-\frac{ \s014 v_*'}{1+(v_*')^2}-\s034  \arctan (v_*')\right]\,,
\eeq
where we have used \eq{eq:full_flow_threshold} for the cutoff shape function $r=r_{\rm opt}$.\footnote{The result \eqref{eq:FP6F} follows after integrating the differential equation $\partial_t v'_*=0$  using the method of characteristics to find $z=z(v'_*)$ from  the flow \eqref{eq:full_flow_threshold} at large $N$. Inverting the family of solutions into $v'_*=v'_*(z)$ and expanding around small fields identifies the free integration constant as $\la{6\rm F}$, thus giving \eqref{eq:FP6F}. For more technical details, see  \cite{Cresswell-Hogg:2022lgg,Cresswell-Hogg:2022lez} and \cite{Litim:2017cnl,Litim:2018pxe}.} All points along the critical line correspond to conformal field theories, connected amongst each other by the exactly marginal perturbation $\sim\la{6\rm F}\int(\bar\psi\psi)^3$. \step

Interestingly, however, the line of fixed points is finite rather than infinite. For $|\la{6\rm F}|\le \la{6\rm F, crit}$, we find a unique and globally well-defined solution $v'(z)$ from \eqref{eq:FP6F} for all values of the fields.\footnote{The numerical values of the critical coupling $\la{4\rm F,*}$ and the critical endpoint $\la{6\rm F, crit}$ are non-universal, reading $-\tfrac{1}{2}$ and $\tfrac{3 \pi}{8}$ for  an optimised fermionic regulator function $r_{\rm opt} ( y ) = ({1}/{\sqrt{y}} - 1 )  \theta \left( 1 - y \right)$ used in \eq{beta2} and \eq{eq:FP6F}. Notice however that their ratio $\la{6\rm F, crit}/(\la{4\rm F,*})^3=-3\pi$ is a {\it universal} number, and independent of the RG scheme and the regulator shape function.} Along the line of fixed points, universal scaling dimensions associated to eigenperturbations $\sim\int(\bar\psi\psi)^n$ are given by $\vartheta_n=n-3$, independent of the marginal coupling $\la{6\rm F}$.
For $|\la{6\rm F}|>\la{6\rm F, crit}$, on the other hand, the fixed point solution no longer corresponds to a well-defined quantum effective action. Exactly at the boundary $|\la{6\rm F}|=\la{6\rm F, crit}$, 
scale symmetry is broken spontaneously and the scaling dimensions are modified. \step

The spontaneous generation of  mass can  be appreciated in terms of a non-perturbative gap equation for the physical fermion mass $M_F=k\,v'|_{z=0}$. For any finite $v'|_{z=0}$, masslessness of the critical theory for $k\to 0$ follows trivially. At the boundary $|\la{6\rm F}|=\la{6\rm F, crit}$, however, the dimensionless mass parameter $v'|_{z=0}$ diverges. To understand the implications, we exploit \eq{eq:full_flow_threshold} to find a  gap equation for the physical fermion mass  $M_F$ at the critical point,
\be
\label{eq:BMB_phenom}
\big[ \la{6\rm F} - \la{6\rm F, crit} \ {\rm sgn}  ( M_F ) \big] M_F = 0\,,
\ee
and in the limit where all fluctuations are integrated out. We observe that as long as the 6F coupling stays below a critical strength $|\la{6\rm F}|<\la{6\rm F, crit}$, the theory remains strictly massless as it should for a conformal field theory.  However, once the critical coupling strength is reached, the gap equation is fulfilled for any $M_F$ (modulo its sign). Hence, quantum scale symmetry is broken spontaneously, leading to the generation of a fermion mass $M_F$ which is not specified by the fundamental parameters of the theory. \step

From the viewpoint of mass generation, the main point here is that fermion mass is generated spontaneously  due to strong fluctuations at a conformal critical point. The key effect is facilitated by {\it dangerously irrelevant} interactions having turned into  exactly marginal ones due to fluctuations. As such, the presence of chirally-odd interactions, and thus the absence of fundamental chiral symmetry,  was a necessity in the first place  to trigger spontaneous scale symmetry breaking, and, consequently, fermion mass is generated without the breaking of any symmetry other than scale symmetry.

\subsection{Critical Yukawa}

Next, we  consider the scalar-Yukawa theory \eqref{eq:SGNY3} in the limit of many fermion flavours  \cite{Cresswell-Hogg:2023hdg}. For suitably chosen microscopic parameters, the theory displays a line of interacting   critical points in the infrared, characterised by the Yukawa coupling $h_*$ and the classically relevant  cubic scalar coupling $\la{3}$ which has become exactly marginal due to fluctuations,
\beq\partial_t\lambda_3\equiv 0\,,\eeq
and which serves as a free parameter. At a fixed point the theory is massless. Once the cubic scalar coupling $\la{3}\neq 0$, chiral symmetry is absent fundamentally. Along the line of fixed points, the quantum effective potential for the scalar field 
$u_*(\sigma)$ can be determined explicitly as a function of $\la{3}$, see \eqref{eq:FPsolutions3}. Most importantly, the line of fixed points is finite rather than infinite, and only for $|\la{3}|\le \la{3, \rm crit}\equiv 3\pi h^3_*$ do we find a unique and well-defined potential  for all values of the fields, including a stable ground state. 
This result  is best appreciated directly from the quantum critical potential. Taking the IR limit $k\to0$  at the fixed point  \eqref{eq:FPsolutions3}, and re-introducing dimensionful fields $\phi\sim k\sigma$ and potential $U(\phi)\sim k^3 u(\phi/k)$, one finds the quantum critical potentials 
\be \label{pot} U_{\rm crit}=\s016\big[ \la{3} + \la{3,\rm crit} \, {\rm sgn} (\phi) \big] \phi^3\,.\ee  
The significance of \eq{pot} is that all parity-even critical scalar self-interactions $\sim \lambda_{2n}\phi^{2n}$ combine into the cusp-like non-analytic term $ \la{3,\rm crit} \, {\rm sgn} (\phi)  \phi^3$.  These contributions are countered by the single term arising form the chirally-odd cubic self-interactions $\la{3}  \phi^3$. Provided  $|\la{3}|<\la{3, \rm crit}$, chirally-even  contributions dominate, leading to a potential which is bounded from below.
If  $|\la{3}|=\la{3, \rm crit}$, however, even and odd contributions cancel out exactly, leading to a potential which is entirely flat for the half-space of positive or negative fields.
Finally, as soon as $|\la{3}|>\la{3, \rm crit}$, the chirally-odd contributions dominate and the potential becomes unbounded from below. This pattern explains why vacuum stability imposes the bound $|\la{3}|\le \la{3, \rm crit}$ on the line of physically acceptable fixed points.\step

The link between the loss of vacuum stability and the spontaneous generation of fermion mass can now be understood as follows. Using \eq{pot} we extract the gap equation
\beq\label{eq:gapGNY}
\big[ \la{3} + \la{3,\rm crit} \, {\rm sgn} \langle \phi\rangle \big] \langle \phi\rangle=0 
\eeq
 for the  ground state of the scalar field. Provided $|\la{3}|\neq \la{3, \rm crit}$, 
 we find that the disordered phase $\langle \phi\rangle=0$ is distinguished, which is the true ground state of \eq{pot} if $|\la{3}|< \la{3, \rm crit}$.
For $|\la{3}|= \la{3, \rm crit}$, however, the potential has a flat direction and the scalar field can take any expectation value in the ordered phase, $\langle \phi\rangle\neq 0$, modulo its sign which correlates with the sign of $\la{3}$. Hence, scale symmetry is broken spontaneously, and the fermions become massive due to residual Yukawa interactions at the fixed point, 
$M_F=h_* \langle \phi\rangle$.
The latter obeys a gap equation analogous to \eqref{eq:gapGNY}
\beq\label{eq:BMB_phenomIR}
\big[ \la{3} + \la{3,\rm crit} \, {\rm sgn} \langle \phi\rangle \big] M_F =0\,,
\eeq
 showing that fermion mass becomes a free, dimensionful parameter at the quantum critical point with $|\la{3}|= \la{3, \rm crit}$, and  unspecified by any of the fundamental parameters of the theory.  The scalar field remains massless. \step
 
 From the viewpoint of mass generation, we conclude that fermion mass is generated spontaneously  at a quantum critical point, and ultimately because strong fluctuations  have turned  a  relevant interaction  into  an exactly marginal one, enabling a degenerate ground state. Crucially, chirally-odd interactions were necessary in the first place  for   scale symmetry breaking to  occur, and fermion mass is generated without the breaking of a discrete symmetry.

\section{\bf Discussion and Conclusions}\label{Sec:C}

We have investigated the generation of fermion mass  in quantum field theories where classically relevant or irrelevant  interactions break  a discrete chiral or parity symmetry. By construction, the setup implies that  mass can be generated explicitly, dynamically, or spontaneously, and without the breaking of a  symmetry. Interestingly, and even though fermion mass is no longer protected by a symmetry, we found that the explicit generation of mass by fluctuations is parametrically suppressed as $1/N$ where $N$ denotes the number of fermion flavours. This  result explains why, at infinite $N$, a vanishing fermion mass at any one scale is  sufficient to protect fermion mass at all scales, and why the more substantial constraint of full chiral or parity symmetry is not required \cite{Cresswell-Hogg:2022lgg,Cresswell-Hogg:2022lez}. In other words,  the  no-longer-by-symmetry-protected fermion mass is nevertheless protected as a consequence of infinitely many degrees of freedom. The  mechanism is appreciated by contrasting \eq{beta1} with \eq{mF=1/N} for Gross-Neveu-type theories, or   \eq{eq:mF2} with \eq{GNY-mF=1/N} for  scalar-Yukawa theories. Hence, at infinite $N$, the  dynamical generation of fermion mass continues to take the form of a second order quantum phase transition, very much like in settings with chiral symmetry, mediated either by strong four- and six-fermion interactions, or  by Yukawas and a non-trivial vacuum expectation value for a scalar field, but without  breaking a  symmetry. On a different tack, we also found that dynamical mass generation in fermionic theories can proceed without an underlying Landau pole, curtesy of  six-fermion interactions (Fig.~\ref{fig:4F} vs.~Fig.~\ref{fig:6F}). \step

What  changes at finite $N$ is that chirally-odd fluctuations do finally generate fermion mass explicitly, even if a fermion mass is absent at the high scale. However, it is important to notice that this effect is  parametrically suppressed \eqref{mF=1/N}, \eqref{GNY-mF=1/N}, and   at least  $1/N$ subleading  to the dynamically-generated   fermion  mass. It follows that the generation of fermion mass, in the absence of chiral symmetry, proceeds through an increasingly  rapid crossover with increasing $N$, which approaches a continuous quantum phase transition  in the limit $1/N\to 0$. Stated differently, the manner in which the  breaking of chiral symmetry by interactions percolates  back into the  generation of  mass is  diluted by the number of fermionic degrees of freedom. What also changes  at finite $N$  is that the operators $\int(\bar\psi\psi)^3$ or $\int \phi^3$ are no longer exactly marginal. Still, these chirally-odd operators  remain near-marginal owing to  strong fluctuations close to a critical point, \eq{la3}, and   continue to contribute  to the dynamically-generated mass, much like at infinite $N$.  We have confirmed that the latter is  invariably enhanced by contributions from chirally-odd interactions, both in Gross-Neveu theories with six-fermion interactions
(Fig.~\ref{4FGN} vs.~Fig.~\ref{6FGN}) and in Gross-Neveu--Yukawa theories with cubic scalar self-interactions (Fig.~\ref{2BGNY} vs.~Fig.~\ref{3BGNY}). It would be interesting to confirm these predictions using Monte-Carlo  simulations on the lattice. \step

In addition, we have  discussed the  spontaneous generation of fermion mass at  quantum critical points. What's different here  is that fermion mass   follows  from the breakdown of quantum scale symmetry, and its value is not determined by any of the fundamental parameters of the theory. We have reviewed two types of critical conformal theories which display this phenomenon, a Gross-Neveu type theory with six-fermion  interactions at an UV critical point \cite{Cresswell-Hogg:2022lgg,Cresswell-Hogg:2022lez}, and a scalar-Yukawa type theory with cubic scalar self-interactions at an IR critical point \cite{Cresswell-Hogg:2023hdg}. In either case, the   primary  driver is that certain chirally-odd interactions, i.e.~the classically irrelevant  operator $\sim \int(\bar\psi\psi)^3$ in the fermionic theory and the classically relevant operator $\sim \int \phi^3$  in the scalar-Yukawa theory, become exactly marginal  quantum mechanically as a result of  strong  residual  fluctuations.
This feature is  used to dial across the finite conformal manifold up to a  point where scale symmetry is broken spontaneously,  \eq{eq:BMB_phenom} and \eq{eq:BMB_phenomIR}, thereby triggering  the  generation of  fermion mass without the breaking  of any   symmetry other than scale symmetry. Clearly,  this    mechanism for fermion mass generation is unavailable  for theories with fundamental chiral symmetry. \step

 Our findings add  to the   studies of models with symmetric mass generation, though with the  twist that  chiral symmetry  has become obsolete due to interactions. While we have focussed on scenarios in 3d theories which could be of relevance for Dirac fermions in condensed matter systems \cite{Herbut:2006cs,Herbut:2009vu,Vafek:2013mpa,Classen:2017hwp} or on the lattice, it will  be interesting to understand  whether similar  mechanisms  may be operative in 4d models of particle physics \cite{Litim:2014uca,Bond:2017tbw,Bond:2019npq,Bond:2021tgu,Litim:2023tym,Steudtner:2024pmd}. This is left for future study.\step\step

{\bf Acknowedgements.} DL thanks Shailesh Chandrasekharan, Anna Hasenfratz, Marina Marin\-kovic, Ethan Neil, Srimoyee Sen, Oliver Witzel, Cenke Xu, Yizhuang You,  and the participants of the workshop {\it Emergent Phenomena of Strongly-Interacting Conformal Field Theories and Beyond} (Aspen, Sept 2023) for discussions. This work is  supported by the Science and Technology Facilities Council (STFC)  under the Consolidated Grant T/X000796/1, and was initiated and performed in part at the Aspen Center for Physics, which is supported by National Science Foundation grant PHY-2210452.

\bibliography{SMGv2}
\bibliographystyle{mystyle}

\end{document}